\documentclass[pre,superscriptaddress,twocolumn,showpacs]{revtex4}
\usepackage{epsfig,amsmath,amssymb,graphicx,color,calc}

\renewcommand{\phi}{\varphi}
\newcommand{\beq}{\begin{equation}}
\newcommand{\eeq}{\end{equation}}
\newcommand{\ba}{\begin{align}}
\newcommand{\ea}{\end{align}}

 \let\b=\beta     
         
          \let\p=\pi  
\let\s=\sigma    \let\ph=\varphi     
           
\let\D=\Delta          
\let\Si=\Sigma         
  \let\th=\theta

\let\io=\infty

\def\to{\rightarrow}
\def\la{\left\langle}
\def\ra{\right\rangle}

\begin{document}

\title{Can the jamming transition be described
using equilibrium statistical mechanics?}

\author{Ludovic Berthier}
\affiliation{Laboratoire des Collo{\"\i}des, Verres
et Nanomat{\'e}riaux, UMR CNRS 5587, Universit{\'e} Montpellier 2,
34095 Montpellier, France}

\author{Hugo Jacquin}
\affiliation{Laboratoire Mati\`ere et Syst\`emes Complexes, UMR CNRS 7057,
Universit\'e Paris Diderot -- Paris 7, 10 rue Alice Domon et L\'eonie 
Duquet, 75205 Paris cedex 13, France}

\author{Francesco Zamponi}
\affiliation{Laboratoire de Physique Th\'eorique, 
\'Ecole Normale Sup\'erieure, 24 Rue Lhomond, 75231 Paris Cedex 05, France}

\date{\today}

\begin{abstract}
When materials such as foams or emulsions are compressed, they 
display solid behaviour above the so-called `jamming' transition. Because 
compression is done out-of-equilibrium in the absence of
thermal fluctuations, jamming appears as  
a new kind of a nonequilibrium phase transition.
In this proceeding  paper, we suggest that tools from 
equilibrium statistical mechanics can in fact be used to 
describe many specific features of the jamming transition.
Our strategy is to introduce thermal fluctuations 
and use statistical mechanics to describe 
the complex phase behaviour of systems of soft repulsive particles, 
before sending temperature to zero at the end of the calculation.   
We show that currently available implementations of standard tools such as 
integral equations, mode-coupling theory, or replica 
calculations all break down at low temperature and large density, but
we suggest that new analytical schemes can be developed to 
provide a fully microscopic, quantitative description of the jamming 
transition. 
\end{abstract}


\maketitle

From the point of view of statistical mechanics, 
the jamming transition observed by compressing 
random packings of soft repulsive particles in the absence 
of thermal fluctuations is an intriguing phenomenon~\cite{vanhecke}. 
It is a phase transition, in the sense that the mechanical 
response of the system changes abruptly at a critical 
density~\cite{durian}. It is also a critical phenomenon, since several 
scaling laws and diverging length scales have been described
on both sides of the transition~\cite{ohernmodel,wyart}. 
Finally, it is a nonequilibrium phenomenon because the transition
occurs in samples prepared out-of-equilibrium in 
the absence of any relevant
thermal fluctuations, and it is thus not possible to describe jamming
without stating precisely the protocol 
used to prepare the system~\cite{KK}. 
Yet, quite remarkably, many features 
of the jamming transition appear to be protocol-independent.  

Given the central role played by far from equilibrium 
critical phenonemona in statistical mechanics~\cite{zwanzig}, 
jamming is thus a very active field of research and attracts the attention
of the statistical mechanics community~\cite{vanhecke,wyartreview}.   
In this conference paper, we mainly discuss the problem of the 
jamming transition from the point of view of statistical mechanics. 
We review, and expand in some places, recently published work 
on the subject, we discuss the philosophy and first results 
of our on-going effort to develop
a new analytical approach to the problem~\cite{tobe}.  

This paper is organized as follows. 
In Sec.~\ref{intro1}, we give a short account of the
properties of the jamming transition, that theory should 
explain and reproduce. In Sec.~\ref{intro2} we briefly review the main
theoretical approaches and explain the basic ideas behind our 
approach.
In Sec.~\ref{fluid}
we show how liquid state theory fares, and explain why it fails 
at low temperature. In Sec.~\ref{glass}, we treat the 
fluid-glass transition of harmonic spheres using both mode-coupling 
theory and replica calculations. We present our conclusions 
in Sec.~\ref{conclusion}. 

\section{Jamming as a nonequilibrium \newline phase transition}
\label{intro1}

In a pioneering study, Durian showed that jamming could 
fruitfully be studied in computer simulations using simple
models of soft repulsive spheres, and he introduced
a model of harmonic spheres interacting through the 
simple pair potential~\cite{durian}
\begin{align}
V(r) = \epsilon 
\left( 1 - \frac r \s \right)^2 \th \left( 1 - \frac r \s \right),
\label{pair_potential}
\end{align}
where $\sigma$ represents the particle diameter, $\epsilon$
is an energy scale, and $\theta(x)$ is the Heaviside function. 
Thus particles repell each other harmonically when they overlap, 
but ignore each other otherwise. 
In Durian's original work, the potential was meant to describe the 
physics of wet foams, but this `bubble' model is in fact so generic that
it could equally be applied to the physics of soft colloids
(such as dense microgels~\cite{vestige}), or emulsions~\cite{mason}. 
Therefore, $\epsilon$ can be
interpreted as a parameter accounting for the elasticity of 
the soft particles, be they soft bubbles, colloids or droplets. 
In the absence of thermal fluctuations or external
forcing, the unique control parameter for the phase behaviour of 
the model of harmonic spheres is the number density, $\rho = N/V$,
for a system composed of $N$ particles enclosed in a volume $V$. 
Equivalently, one can use the `packing fraction', 
$\phi = \pi \sigma^3 \rho / 6$, although this name only 
strictly makes sense when particles do not overlap at low enough density.
Indeed, because particles are soft, $\phi$ is not bounded.

 \begin{figure}
\includegraphics[width=8.5cm]{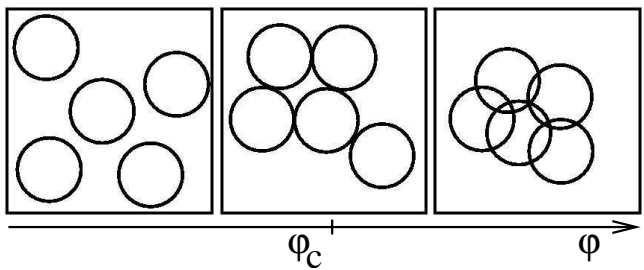}
\caption{Packings of soft repulsive spheres 
at $T=0$ undergo a nonequilibrium jamming 
transition at a critical packing fraction $\phi_c$.
The mechanical properties change from fluid to solid 
across $\phi_c$. Energy, pressure, and mechanical 
moduli increase continuously from zero above $\phi_c$, while the number
of contact per particle jumps discontinously to a finite value 
at $\phi_c$.}
\label{fig_fig}
\end{figure}
 
In Fig.~\ref{fig_fig} we summarize schematically the main features of the 
jamming transition that have been discovered, mostly
through numerical observations~\cite{vanhecke}. Simulations
have revealed the existence of a critical packing fraction, 
$\phi_c$, below which the packings have no overlap. In this 
phase, the energy, pressure and number of contacts between particles 
are zero.
Above $\phi_c$, pressure, energy, number of contacts, bulk and shear
moduli are non-zero. The linear (elastic) response of the system 
to deformation changes abruptly at the transition: 
below $\phi_c$, small deformations can be
imposed without any energy cost, while above $\phi_c$ an arbitrarily 
small deformation causes an increase in energy.

Interestingly, almost all these quantities increase continuously 
from zero across $\phi_c$
and vary algebraically with the distance to the transition, 
$\phi - \phi_c$. A notable exception is the number of contacts, 
which, as suggested by Fig.~\ref{fig_fig}, jumps discontinuously 
to a finite value at $\phi_c$ given by $z_c = 2d$ where $d$
is the space dimensionality. The critical $z_c$ corresponds 
to the ``isostatic'' value, i.e. the minimal value for the system
to be rigid~\cite{alexander}. At $\phi_c$ the system is therefore marginally
solid, with vanishing shear and bulk moduli. Moreover, 
the pair correlation $g(r)$ of systems at $\phi_c$ is very different
from both amorphous glasses or dense liquids, and present 
a number of singular behaviours, from a diverging contribution
at the interparticle distance $r=\sigma$~\cite{donev,silbert}, to 
large-distance anomalies~\cite{torquato} reflected in a peculiar 
low-wavevector behaviour of the isothermal compressibility~\cite{peter}.

Finally, the nonequilibrium nature of the transition is clear 
from the fact that thermal fluctuations are actually irrelevant 
for the features described above. This implies that `crossing' the 
transition by `compressing' the system is not a 
uniquely-defined procedure~\cite{commenttorquato}. It is thus crucial 
to also specify how this is done in practice, in particular how
averages at each density are performed, that is, how 
different configurations at the same density are sampled.
In an equilibrium setting, this is not necessary as
configurations are sampled with their 
associated Boltzmann weights.  Two well-studied procedures
to study the jamming transition are the following.
First, rapid compressions of hard sphere systems simulated by 
molecular dynamics~\cite{LS} produce jammed configurations in the limit 
of infinite pressure~\cite{welldefined,PZ}. 
Note that in this procedure thermal 
fluctuations play a role since thermal equilibrium can be reached
at low enough density. Note also that 
the jammed phase cannot be accessed since particles cannot overlap.
A second and very different procedure consists of studying systems of 
soft repulsive spheres, as in Eq.~(\ref{pair_potential}), directly 
at $T=0$ using energy minimization methods~\cite{ohernmodel}. 
In both cases, 
sampling and averages are performed by repeating the 
compression or minimization protocol from a different set of initial 
conditions. A critical volume fraction 
$\phi_c$ with the properties described above is 
generically found for any of these protocols. At present,
there is numerical evidence that if the exact value of 
$\phi_c$ is protocol dependent, the critical properties 
are not~\cite{jline}.

\section{On the theory of the jamming transition}
\label{intro2}

At the theoretical level, there is at present no 
accepted framework to understand and account for all the
features of the jamming transition that we have described, 
which mostly stem from very detailed numerical observations~\cite{vanhecke}. 
Thus, if the physics and the main features of the
transition are well described~\cite{wyartreview}, the theory is by comparison 
less advanced. An elegant scaling theory, that identifies a
divergent length scale and relates various observed
scaling laws or critical exponents, was elaborated~\cite{wyart,wyartreview}.
Additionally, several distinct statistical frameworks 
were also developed~\cite{makse,maxime,henkes},
using in particular the idea that 
a statistical ensemble (the `Edwards ensemble'~\cite{edwards} or a `force
ensemble'~\cite{ellenbroek}) 
distinct from the Gibbs ensemble must be introduced to study 
the jamming transition. A detailed account of these approaches
is behind the scope of this short paper and we refer the reader to
the recent reviews~\cite{vanhecke,wyartreview} for further references.

Here, we shall argue that the three main characteristics
discussed above, namely the existence of a fluid-solid 
jamming phase transition, 
its associated critical properties, and its nonequilibrium nature can 
in principle all be accurately computed within the standard framework 
of equilibrium statistical mechanics, starting from the sole knowledge
of the interaction between the particles in 
Eq.~(\ref{pair_potential}). The general strategy we propose is to study 
the statistical mechanics of a system of harmonic spheres 
in the presence of thermal fluctuations. If this procedure 
is properly implemented, then we expect that a sharp phase transition 
with the above characteristics will occur at a critical volume fraction in 
the limit of $T \to 0$. 

The main conceptual difficulty to be faced has actually a physical 
origin, and deep consequences. In the relevant regime, the 
system is in fact characterized by the existence of a large 
number of amorphous metastable states separated by large free energy
barriers~\cite{KK}. This means that the system gets naturally
dynamically arrested in nonequilibrium glassy states, and that the 
free energy landscape is quite complex. This implies first
that the above mentioned
dynamical protocols to probe the jamming transition are naturally
affected by the glass transition~\cite{KK,PZ}. Second, this means that 
statistical mechanics treatments must accurately 
take the complexity of the phase space into 
account---or will necessarily fail.
A theoretical framework to handle this complexity was developed
in Ref.~\cite{monasson} and applied in the context
of particle systems in Refs.~\cite{MP96,MP2,MP}. 
It was also more recently 
implemented in Ref.~\cite{PZ} for hard spheres.

The aim of the present work is to show that, 
although the approach is conceptually simple and direct, there remain 
purely technical difficulties, since previous implementations of these ideas
are not very accurate in the limit of 
interest to describe the jamming transition, 
where temperature is low and density is large. 
In a separate work~\cite{tobe}, we describe our ongoing efforts to 
circumvene these technical difficulties to try and 
yield a predictive, quantitative, microscopic 
approach to the jamming transition.  

\section{Statistical mechanics of harmonic spheres: fluid phase}

\label{fluid}

\subsection{Integral equations} 

\label{anomaly}

Our first step is to start investigating the behaviour of
repulsive harmonic spheres in a regime where temperature is large.
In the absence of attractive forces, the system is thus in a fluid
phase, where correlations between particles are small. 
  
In this regime, it is enough to treat the system using 
tools developed to study the statistical mechanics of the 
liquid state, such as integral equations~\cite{hansen}. These are 
typically closure relations yielding the pair correlation
function of the fluid, 
\beq
g(r) = \frac{1}{\rho N} \la \sum_{i \neq j} \delta 
( |{\bf r} - {\bf r}_{ij} |) \ra,
\eeq
where brackets denote a thermal average.
 
In Ref.~\cite{hugoSM} we studied the potential (\ref{pair_potential}) using 
one particular closure relation, the Hyper-Netted Chain (HNC) approximation.
In this approximation, 
\beq
g(r) = \exp[ -\beta V(r) + g(r) -1 - c(r) ],
\label{hnc}
\eeq
where $\beta = 1/T$,
$c(r)$ is the direct correlation function defined 
through the Ornstein-Zernike equation:
\beq
g(r)-1 = c(r) + \rho \int d{\bf r}' c(|{\bf r}-{\bf r}'|) [g(r')-1].
\label{OZ}
\eeq
By numerical integration of Eq.~(\ref{hnc}),
one can get $g(r)$, and thus study the thermodynamic behaviour 
of the model for any state point $(\phi,T)$. 

Not surprisingly, no phase transition is detected 
by compressing the system at constant temperature within 
such an approach, even when temperature is small. In fact, it can
easily be shown that the ground state energy density, 
\beq
e_{\rm gs}(\phi) = \lim_{T \to 0} e(T,\phi)
\eeq 
remains zero at all $\phi$, since the energy density
vanishes as $e (T,\phi) \sim T^{3/2}$, a result 
which is clearly inconsistent with the cartoon in Fig.~\ref{fig_fig}. 
Simultaneously,
the pair correlation function remains smooth and shows none of the 
jamming singularities described in the introduction in the range 
where the jamming transition is found numerically. 
As we argue below, this is not an artefact of the specific 
approximation employed, but is likely a generic feature 
of integral equations developed for liquids.

\begin{figure}
\includegraphics[width=8.5cm]{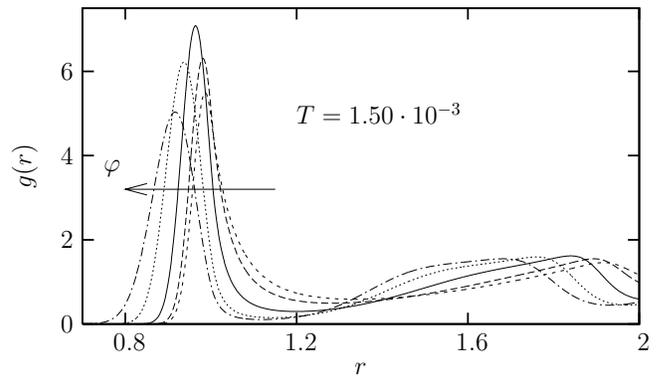}
\caption{Nonmonotonic evolution of the pair correlation function
when volume fraction is increased at constant temperature
in the fluid phase, as predicted using the hypernetted chain 
approximation~\cite{hugoSM}, 
a simple integral equation frequently used in
liquid state theory. Volume fractions are 
$\varphi=0.600$, 0.663, 0.777, 0.900, and 1.00 (from right to left).
}
\label{fig_vestige}
\end{figure}
 
Although a sharp jamming transition is not found, 
there is an interesting feature which emerges from the study 
of integral equations, as shown in Fig.~\ref{fig_vestige}.
Upon compression at constant temperature, the evolution of the
first and second peaks in the pair correlation function  
shows two distinct regimes, depending on the value of the 
volume fraction. For $\phi < \phi^\star(T)$, 
the position of the first peak of $g(r)$ shifts to smaller distances,
reflecting the fact that particles get closer to each other. 
The height of the peak increases and structural order
increases in the fluid. For $\phi > \phi^\star$, the position 
of the peak continues to shift to smaller $r$, but the height of 
the peak now decreases with $\phi$.  Thus, the fluid becomes more
disordered as density is increased, an `anomalous' 
behaviour not seen in simple liquids.

This density anomaly is actually well-known in the context 
of the physics of ultrasoft particles~\cite{ard,likos}, 
i.e. particles interacting 
with a pair potential that remains finite when particles fully overlap,
as in Eq.~(\ref{pair_potential}) where $V(r=0) = \epsilon$, but it is
interesting to rephrase its physical explanation in the 
context of jamming~\cite{hugoSM}, where the crossover 
$\phi^\star$ plays the role of a `soft jamming'~\cite{ken} or a `thermal
vestige'~\cite{vestige} 
of the jamming transition. It is therefore also particularly 
interesting in the context of the question asked in our title, 
since the physics behind the density anomaly is a
competition between energy and entropy to minimize the free energy 
at thermal equilibrium. Briefly, the anomaly stems from 
the increasing difficulty, and thus the increasing entropic cost, 
to find states with little overlap, and thus with little energy, 
when $\phi$ becomes large. Above $\phi^\star$, 
the system thus prefers instead paying some 
finite energy to allow particle overlaps, which can be done 
in many different ways and thus makes the structure of the 
fluid less ordered. 

As shown in Fig.~\ref{fig_fig}, the jamming transition corresponds
to an extreme case of the crossover observed at finite 
temperature, namely a transition betwen a low-$\phi$ phase
were numerous states with no overlap can be found, and 
a large-$\phi$ phase where states without overlap only exist with 
a vanishing probability. This suggests theories of the liquid 
state fail to describe a sharp jamming transition 
because they do not accurately describe this competition at very low
temperatures and large densities, as we confirm in the following
section.

\subsection{Numerical simulations}

\begin{figure}
\includegraphics[width=8.5cm]{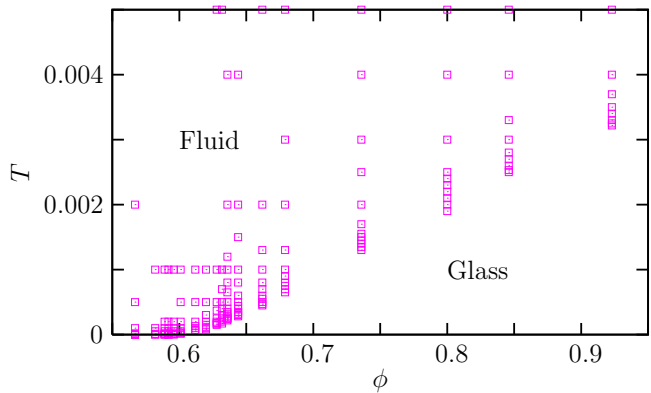}
\caption{Numerical phase diagram for a binary mixture 
of harmonic spheres~\cite{tom2}. 
The symbols correspond to `fluid' state points 
where thermal equilibrium can be reached using molecular dynamics
simulations. In the `glass' phase with no symbols, 
the relaxation time has become too large to be numerically determined,
the system is effectively a nonergodic glass. Note that the jamming 
transition at $T=0$ and $\phi=\phi_c$ cannot be accessed at thermal
equilibrium. Fluids do not jam, only glasses do.}
\label{fig_phase}
\end{figure}

Why does liquid state theory fail to account for 
the jamming transition? A direct answer is provided by numerical
studies of the finite temperature behaviour of the model
(\ref{pair_potential}). In Refs.~\cite{tom,tom2} the dynamics
of harmonic spheres was studied using molecular dynamics
simulations. In Fig.~\ref{fig_phase} we report in 
the $(\phi,T)$ plane the state points for which thermal equilibrium
could be reached, and the typical relaxation time measured 
with sufficient precision during the course of the numerical study.
Numerical simulations thus indicate that it becomes 
increasingly difficult to reach equilibrium in the regime 
where density is large and temperature low, see Fig.~\ref{fig_phase}.
In the `glass' phase, structural relaxation does not occur
during the timescale allowed by the numerical experiment, and the
system is essentially frozen in a very long-lived metastable state.
For practical purposes, it has all the characterisitcs of 
a glass~\cite{glassreview}, 
i.e. an amorphous (liquid-like) structure, which does 
not relax (solid-like) on the observation timescale. 

Therefore, simulations teach us that the free energy landscape
of the system of harmonic spheres becomes very complicated 
in the glass regime~\cite{KK}. It is this complexity which is responsible
for the breakdown of liquid state theory, which cannot be used 
to explore the glass phase. 

In Fig.~\ref{fig_phase} we also note that along the $T=0$ axis, 
equilibration cannot easily be achieved above 
$\phi \approx 0.60$, while most numerical 
determinations of $\phi_c$ are much above this value, in the range
$\phi_c \approx 0.64-0.66$.
Thus, we conclude that even if one introduces 
thermal fluctuations into the game, the jamming transition 
cannot be crossed at thermal equilibrium because the glass transition
intervenes first, whatever the path in the $(\phi,T)$ plane which 
is followed, compressions or quenches. 
Therefore, the jamming transition can only be observed 
by compressing glasses---not fluids. 

The unavoidable conclusion
is that a theory of the glass state is needed, instead 
of a theory of the liquid state, as we describe 
in the next section.

\section{Statistical mechanics of harmonic spheres: glass phase}

\label{glass}

Although the theoretical literature of the
glass transition is vast~\cite{glassreview}, there are not very many 
microscopic quantitative approaches, i.e. capable
of formulating quantitative predictions starting from 
the knowledge of the interaction between the particles, 
as we attempt to do here. The mode-coupling theory of the 
glass transition and the replica approach to the glass phase 
are the two examples we discuss in this section. 

\subsection{Mode-coupling theory}
  
The mode-coupling theory of the glass transition was developed
in the mid-80's~\cite{gotze}. It is built using the tools first developed 
to describe the structure and dynamics of liquids, and, 
in its initial formulation, 
uses the formalism of projection operators to derive a closed 
set of dynamical equations of motion for time correlation 
functions of supercooled liquids. In its common implementation
mode-coupling theory can thus be seen as a `black box' which 
is fed by structural information on the fluid (the two-point 
static structure factor of density fluctuations), and provides
as an outcome the time dependence of density-density autocorrelation 
functions at any wavevector.

\begin{figure}
\includegraphics[width=8.5cm]{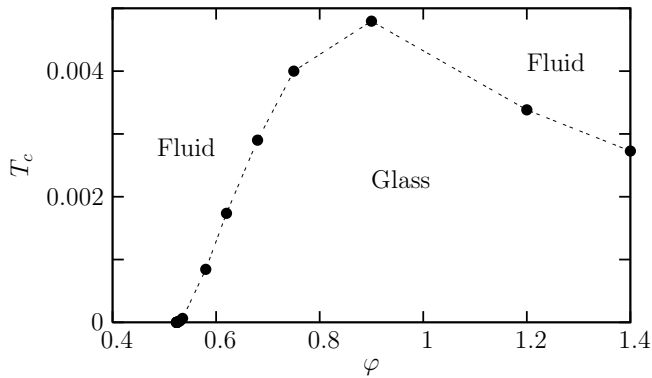}
\caption{Dynamical phase diagram for harmonic spheres derived from 
mode-coupling theory~\cite{hugoPRE} 
confirms the propensity of harmonic spheres
to undergo a glass transition at low temperatures. The reentrant glass line
at large $\phi$ is a dynamical counterpart of the density anomaly
shown in Fig.~\ref{fig_vestige}.}
\label{fig_mct}
\end{figure}

This short description shows that mode-coupling theory is in principle
able to describe the features revealed by the numerical results 
shown in Fig.~\ref{fig_phase}. Indeed we present in Fig.~\ref{fig_mct}
results from the analytical study of the phase diagram of 
harmonic spheres. Part of these results were presented in  
Ref.~\cite{hugoPRE}, where all details can be found. In this 
work, a detailed analysis of the dynamic scaling properties 
near the fluid-glass transition occurring near 
$T=0$ was presented. In Ref.~\cite{angel} a similar theoretical analysis was 
performed for an interaction potential very close to the harmonic 
potential studied here, where the harmonic exponent $2$ 
in Eq.~(\ref{pair_potential}) is changed to $5/2$, the so-called 
Hertzian sphere potential, yielding results qualitatively 
similar to those presented in Fig.~\ref{fig_mct}. 

Importantly the results of the mode-coupling analysis shown 
in Fig.~\ref{fig_mct} confirm the propensity of the harmonic
sphere system to form a glass at large density and low temperature. However,
the mode-coupling results are quantitatively not
very accurate, as is well-known after two decades of mode-coupling 
studies in the field of the glass transition~\cite{glassreview,gotze}.

Although quantitative agreement is not found between theory
and simulations, mode-coupling theory makes one striking prediction
for the phase behaviour shown in Fig.~\ref{fig_mct}. Theory predicts
the existence of a temperature regime where the fluid
of harmonic spheres becomes a solid glass upon compression at constant 
temperature. Strikingly, upon compressing the glass further, 
there exists a second critical volume fractions above which the glass
is melted and becomes a flowing fluid.  Such a glass melting 
at high density is typically not observed in standard 
models of liquids, and is the direct result of the 
particle softness~\cite{angel}. We in fact already provided an explanation 
for this behaviour when we discussed the physics of the density
anomaly in Sec.~\ref{anomaly} and Fig.~\ref{fig_vestige}.
The reentrant glass transition predicted theoretically 
in Fig.~\ref{fig_mct} had not been observed in Refs.~\cite{tom,tom2}, 
because the range of volume fractions studied numerically
was too narrow, but it was recently observed for a system of Hertzian 
spheres in Ref.~\cite{angel}.

Now, although the mode-coupling equations can in principle be used 
to describe the (nonergodic) dynamics in the glass phase, it needs 
as input the static structure of the fluid, which is the very problem
we ascribed ourselves at the beginning of the paper. Thus, 
although this approach is useful in determining the existence and location
of the glass phase in the system, it can not be employed to describe the 
thermodynamic behaviour and structural evolution 
deep into the glass phase. To do so, 
we must turn to the thermodynamic replica approach.

\subsection{Thermodynamic replica approach}

\label{replica}

\subsubsection{The replica method}

The replica approach to the glass transition~\cite{monasson,MP96,MP2,MP} 
can be seen as a modern and more systematic 
implementation of the self-consistent phonon theory
developed long ago by Wolynes and coworkers~\cite{wolynes}.
The goal is to properly evaluate the thermodynamics
of the glass phase, by carefully taking into account the existence
and proliferation of long-lived metastable states which are indeed
responsible for the breakdown of liquid state approaches at low 
temperatures, as described in Sec.~\ref{fluid}. 

To account for the effect of metastable states, 
the partition function is decomposed into the respective 
contributions of inter- and intra-states free energies,
\beq
Z = \int df e^{-N \beta (f - T \Sigma(f))},
\label{sigma}
\eeq  
such that ${\cal N}(f) = \exp[N \Sigma(f)]$ 
represents the number of metastable states with free energy $f$,
and $\Si(f)$ is called `complexity'.
This decomposition can be seen as a generalization of the one based
on the concept of inherent structures developed earlier 
by Goldstein~\cite{goldstein} and Stillinger and Weber~\cite{SW82}.

The replica approach is a computational tool devised to derive 
analytically the complexity of the system as it approaches
the glass transition, and the thermodynamic properties
of the system deep in the glass phase~\cite{monasson,MP}. 
It is based on the introduction of an effective temperature
$T_{\rm eff}= T/m$ conjugated to the free energy $f$ in Eq.~(\ref{sigma}),
in the same way as temperature is conjugated to energy in 
standard computations. If $m$ is an integer, it can be
interpreted as a number of replicas of the original system~\cite{monasson}. 
A careful analysis shows that one is then able
to deduce the thermodynamics of the glass from the thermodynamics of 
a $m$-times
replicated liquid, with $m$ being analytically continued to non-integer values.
To compute the properties of the glass, one must then 
study the properties of an effective liquid which is 
a `mixture' of $m$ copies of the original system~\cite{MP96,MP}.
We refer to Refs.~\cite{giorgio,PZ} for 
extensive reviews of a technique whose advantages and shortcomings, 
successes and failures, are by now well-established.  

As briefly mentioned in the introduction, this 
step is of course crucial to understand the physics 
of the glass transition, but it is equally fundamental 
to properly describe the jamming transition which lies deep into the 
glass phase. The numerical 
protocols devised to study the jamming transition indeed all rely 
on a dynamical exploration of the ground state properties of the system
at $T=0$ found either using gradient descent methods (which indeed produce
inherent structures~\cite{SW82}), or rapid
compressions. Given the topology of the phase 
diagram in Fig.~\ref{fig_phase}, {\it both rapid quenches 
and fast compressions hit the glass transition} at some point, 
and from that point these numerical procedures in fact {\it follow the 
zero temperature properties of long-lived metastable states across 
the jamming transition}.
The thermodynamic replica method, centered around the properties
of these states, appears therefore very well-suited to describe the jamming
transition in soft repulsive systems.

As discussed at length in the glass literature~\cite{glassreview,giorgio}, 
the decomposition
in Eq.~(\ref{sigma}) relies on the existence of 
infinitely long-lived states, such that a thermodynamic
calculation makes sense. This is a typical {\it mean-field}
assumption, because in finite dimensions metastable states only have a finite
lifetime (which of course becomes 
extremely large in the glass phase), and this makes the thermodynamic
replica approach prone to criticisms.

Approaching the jamming transition, though, the lifetime of the states 
indeed diverges, 
so that we expect the mean-field approximation to behave better
 upon approaching the jamming point than it does around the glass 
transition~\cite{PZ}.
We conclude therefore that
thermodynamic replica calculations offer a promising theoretical framework 
to study the ground state properties of harmonic sphere glasses.  
 
\subsubsection{First attempt: Replicated HNC}

In practice, the thermodynamics of the replicated liquid has to 
be computed using
some liquid theory approximation.
The simplest of these approximations is the replicated HNC theory 
developed in Ref.~\cite{MP96}.
It leads to a set of coupled equations for the diagonal and off-diagonal 
replica correlation functions, whose solution allows in principle to 
determine the full phase diagram of the system. 
However, it has been checked in the case of hard spheres that 
this approximation is correct in the
liquid phase, while it fails badly in the glass phase and in particular close
to jamming~\cite{MP96}. The origin of this failure has been discussed 
in \cite[section IV]{PZ},
where the replicated HNC equations are 
presented in full detail.

Therefore, this approximation can only be used to determine
the glass transition lines $T_K(\ph)$ and $T_d(\ph)$
for harmonic spheres. Within mean-field theory, 
the former represents the thermodynamic glass transition where
the complexity vanishes, while the second 
represents the dynamic glass transition marked by the appearance
of metastable states (it is closely related, 
in principle, to the mode-coupling singularity discussed above).

\begin{figure}
\includegraphics[width=8.5cm]{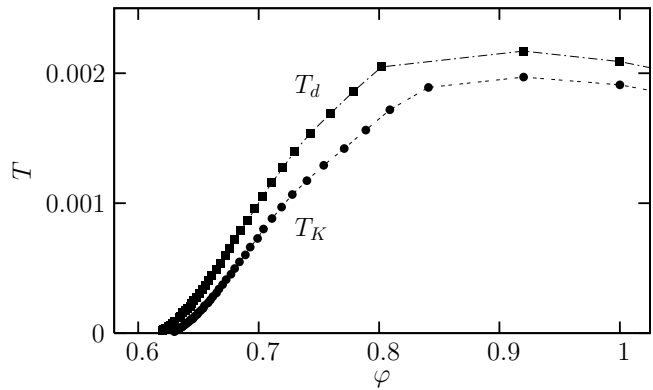}
\caption{Thermodynamic, $T_K(\ph)$, and dynamic, $T_d(\ph)$, 
glass lines determined from solving the replicated HNC approximation
are in good qualitative agreement with both simulations, Fig.~\ref{fig_phase},
and mode-coupling results, Fig.~\ref{fig_mct}.
}
\label{fig_HNCrep}
\end{figure}

These new results are shown in Fig.~\ref{fig_HNCrep}. The curves have the 
expected shape and suggest, once again, the existence 
of a glass phase at low temperatures and large density, 
as seen in computer simulations, see Fig.~\ref{fig_phase}.
In particular, $T_d(\ph)$ has the same qualitative behavior of the 
mode-coupling transition
temperature in Fig.~\ref{fig_mct}, suggesting that
the reentrant glass transition scenario should be a robust feature 
of this system~\cite{angel}.
Note also that the ratio $(T_d-T_K)/T_K$ is often (inversely) correlated
to the kinetic fragility of glassforming systems. 
Then, one prediction of the replicated HNC approach is that this 
ratio decreases on increasing density
above $\ph \sim 0.6$, which is consistent with the 
numerical observation that  the liquid becomes more 
fragile when $\phi$ increases~\cite{tom,tom2}.

\subsubsection{Second attempt: The small cage expansion}

A more successful approach was later suggested 
by M\'ezard and Parisi, based on the physical idea that deep 
into glassy phases when temperature is small, the $m$ replicas
of the effective system will be undergo small vibrations within 
each state, suggesting that a perturbative expansion of the free energy
in the small cage size, $A$, can be performed.  

With this procedure, the free-energy of the replicated system is mapped 
onto the free-energy of a non-replicated system  at 
effective temperature $T_{\rm eff}=T/m$. The glass phase is found to be 
described by values of $m$ that are smaller than $1$, 
so that the replicated liquid is found, in the small
cage expansion, to be equivalent to a non-replicated liquid at higher 
temperatures. As a result of this purely analytical game, 
one can finally relate the thermodynamics of a state point of the original 
system located inside the glass phase, to those of a 
state point in the liquid phase of the effective system, where 
liquid state theory, such as HNC, can reliably be applied.
This scheme has been successfully applied to Lennard-Jones glasses
in~\cite{MP,MP2}.

\begin{figure}
\includegraphics[width=8.5cm]{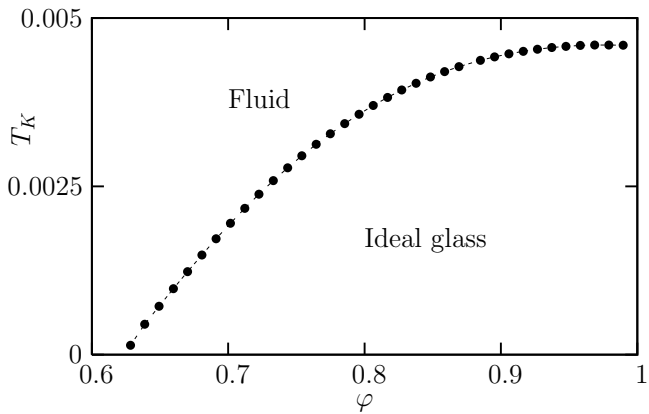}
\caption{The ideal glass transition line, 
$T_K(\ph)$, determined using the first order small cage expansion 
of the replica approach. Note that no transition is found below 
$T_K \sim 1.4 \cdot 10^{-4}$, and the $T \to 0$ limit cannot be 
studied.}
\label{fig_small}
\end{figure}

We have studied the system of harmonic spheres using the 
small cage expansion at first order.
From the replicated free-energy, 
one can deduce the locus of the Kauzmann thermodynamic glass 
transition, and all thermodynamic properties of the glass. 
We present our results for $T_K(\varphi)$ in Fig.~\ref{fig_small}, 
which is again in qualitative agreements with other schemes described
above.

Unfortunately, the small cage 
expansion, while being well-defined 
in the high density regime, fails in the jamming regime that 
we are interested in. The reason for this failure is that the 
non-analyticity at $r=1$ 
of the harmonic spheres (which is mandatory for them to behave like hard 
spheres when $T=0$) makes the free energy expansion 
ill-behaved in the limit $T \to 0$ when density is too small. 
This can be seen in Fig.~\ref{fig_small}, where the glass transition 
line cannot be followed down to $T_K = 0$, and the transition
abruptly disappears at some volume fraction, which is an artefact 
of the small cage expansion.

For the learned readers,
this can be easily seen by looking at the first order expression 
given in Ref.~\cite{MP}.
The first correction to the free energy at small $A$ has the form $-\b \Delta 
F=- 12 A \b \ph (m-1) \int_0^{\io} r^2 g(r) \D V(r) dr$. We 
know that $g(r) \sim e^{-\b V(r)}$ as $T \to 0$. We then 
deduce that $-\b \D F \sim 36 A \sqrt{\p \b} \frac{1-m}{\sqrt{m}}$.
Since the zero-th order term of $-\b F$ has a finite limit (the entropy
of hard spheres), we conclude that the expansion is not defined in 
the limit $T\to 0$, because of the $A \sqrt{\beta}$ prefactor, 
and thus cannot be used to study the approach to the jamming 
transition from above.

This failure also explains why an alternative free energy 
expansion was recently developed to study hard spheres~\cite{PZ}.
In this approach, a different small cage expansion was 
performed in powers of  $\sqrt{A}$ (instead of $A$ in the 
M\'ezard-Parisi scheme), consistent with the observation 
that the $A$-expansion is divergent. Using the 
$\sqrt{A}$ expansion, it was possible 
to study the jamming transition (pressure, structure)
on the hard sphere side~\cite{PZ}. However, this method 
only applies to hard spheres and it cannot be used 
to explore the jamming transition of harmonic spheres.

Technically, a full description of the jamming transition thus
requires an expansion scheme which is able to describe the crossover
between the $\sqrt{A}$ and the $A \sqrt{\beta}$ expansions of the free energy,
valid on both sides of the jamming transition. 
Such an new approximation forms the core 
of a separate publication~\cite{tobe}.

\section{Conclusion}

\label{conclusion}

In this proceeding paper, we have justified  
our on-going effort to attack the 
purely geometric problem of soft sphere packing
in three dimensions using the tools of equilibrium 
statistical mechanics. This philosophy thus suggests to add
a temperature axis to the phase diagram and study the 
statistical mechanics of soft repulsive spheres, the jamming 
transition being obtained in the limit of $T \to 0$. 
Adding thermal fluctuations to a situation where hard constraints 
need to be satisfied is actually a common tool, for instance in combinatorial 
optimization problems~\cite{KK,optimization}. 

However, we also showed that
adding temperature does not immediately solve
the problem, since one quickly realizes that jamming does not 
occur in a fluid but in the 
glassy part of the phase diagram where thermal equilibration is 
not easily reached. Thus, describing jamming requires the development 
of accurate analytical tools to describe the structure and thermodynamics
of soft repulsive glasses, which is a delicate task.

We showed that although conceptually 
feasible, previously published analytical schemes actually fail  
near the jamming singularity, suggesting that new theoretical 
developments are needed to succesfully derive a fully microscopic, 
quantitative theory of the jamming transition. We shall report 
elsewehere the results of our work in this direction~\cite{tobe}.

\end{document}